\documentclass[12pt,a4paper]{article}
\usepackage{graphics,epsfig,rotating,amssymb}
\begin{document}
\newcommand{\goo}{\,\raisebox{-.5ex}{$\stackrel{>}{\scriptstyle\sim}$}\,}
\newcommand{\loo}{\,\raisebox{-.5ex}{$\stackrel{<}{\scriptstyle\sim}$}\,}

\begin{center}
{\large \bf Statistical description of nuclear break-up.}
\end{center}
\vspace{0.5cm}
\begin{center}
{\Large A.S.~Botvina$^{a,b}$ and I.N.~Mishustin$^{b,c}$}\\
\end{center}
\begin{center}
{\it
 $^a$Institute for Nuclear Research, Russian Academy of Sciences, 117312 Moscow,
 Russia\\
 $^b$Frankfurt Institute for Advanced Studies, J.W. Goethe University, D-60438
     Frankfurt am Main, Germany\\
 $^c$Kurchatov Institute, Russian Research Center, 123182 Moscow, Russia\\
}
\end{center}
\normalsize

\vspace{0.3cm}
\begin{abstract}
We present an overview of concepts and results obtained with 
statistical models in study of nuclear multifragmentation. 
Conceptual differences between statistical and dynamical approaches, 
and selection of experimental observables for identification of 
these processes, are outlined. 
New and perspective developments, like inclusion of in-medium 
modifications of the properties of hot primary fragments, are 
discussed. We list important applications of statistical 
multifragmentation in other fields of research. 
\end{abstract}

\vspace{0.2cm}


\vspace{0.5cm}

{\bf 1. Introduction}\\

Statistical models have proved to be very successful in nuclear
physics. They are used for description of nuclear decay when an equilibrated 
source can be identified in the reaction. The most famous example of such a
source is the 'compound nucleus' introduced by Niels Bohr in 1936 
\cite{Bohr}. It was clearly seen in low-energy nuclear reactions leading to 
excitation energies of a few tens of MeV. It is remarkable that this 
concept works also for nuclear reactions induced by particles 
and ions of intermediate and high energies, when nuclei break-up into many 
fragments (multifragmentation). 
According to the statistical hypothesis, initial dynamical 
interactions between nucleons lead to re-distribution of the available 
energy among many degrees of freedom, and the nuclear system 
evolves towards equilibrium. 
In the most general consideration the process may be 
subdivided into several stages: (1) a dynamical stage leading to 
formation of equilibrated nuclear system, (2) disassembly of the system 
into individual primary fragments, (3) deexcitation of hot primary 
fragments. Below we consider these stages step by step. 
In this paper we give an overview of main results obtained 
with statistical models in multifragmentation studies, and analyze 
the most important problems (see also reviews \cite{Gross,SMM,DasGupta1}). 
Several hundred papers concerning multifragmentation were published during 
last two decades, and we apologize that in a short review we can not mention 
all works related to this field. 

\vspace{0.5cm}

{\bf 2. Formation of a thermalized nuclear system}\\

At present, a number of dynamical models is used for description of 
nuclear reactions at intermediate energies. The Intranuclear Cascade Model 
was the first one used for realistic calculations of ensembles of 
highly excited residual nuclei which undergo multifragmentation, see e.g. 
\cite{Botvina90}. Other more sophisticated models were also used for 
dynamical simulations of heavy-ion reactions, such as quantum 
molecular dynamics (QMD), Boltzmann (Vlasov)-Uehling-Uhlenbeck (BUU, VUU) 
and other similar models (see e.g. refs. \cite{dynmodels}).
All dynamical models agree that the character of the dynamical 
evolution changes after a few rescatterings of incident nucleons, when high 
energy particles ('participants') leave the system. 
This can be seen from distributions of nucleon velocities and density 
profiles in remaining spectators \cite{buusmm,Bondorf94,larionov,NMD}. 
However, the time needed for equilibration and transition to the 
statistical description is still under debate. This time is estimated 
around or less than 100 fm/c for spectator matter, however, it slightly 
varies in different models. Apparently, this time should be shorter for 
participant zone produced in heavy-ion collisions at energies above 
the Fermi energy, as a result of initial compression. Parameters of the 
predicted equilibrated sources, i.e. their excitation energies, mass 
numbers and charges vary significantly with this time. 
We believe that the best strategy is to use results of the dynamical 
simulations as a qualitative guide line, but extract parameters of 
thermalized sources from the analysis of experimental data. 
In this case, one can avoid uncertainties of dynamical models in describing 
thermalization processes.

\vspace{0.7cm}

{\bf 3. Break-up of a thermalized system into hot primary fragments} 

\vspace{0.7cm}

{\bf 3.1 Evolution from sequential decay to simultaneous break-up.} \\

After dynamical formation of a thermalized source, its further evolution 
depends crucially on the excitation energy and mass number. 
The standard compound nucleus picture is valid only at low excitation
energies when sequential evaporation of light particles and fission are 
the dominant decay channels. Some modifications of the evaporation/fission 
approach were proposed in order to include emission of fragments heavier 
than $\alpha$-particles, see e.g. \cite{Moretto,Botvina87,charity0}. 
However, the concept of the compound nucleus cannot be applied at high 
excitation energies, $E^* \goo$ 3 MeV/nucleon. The reason is that the time 
intervals between subsequent fragment emissions, estimated both within the 
evaporation models \cite{charity} and from experimental data \cite{jandel}, 
become very short, of order of a few tens of fm/c. In this case there will 
be not enough time for the residual nucleus to reach equilibrium between 
subsequent emissions. 
Moreover, the produced fragments will be in the vicinity of each other and, 
therefore, should interact strongly. The rates of the particle emission 
calculated as for an isolated compound nucleus will not be reliable 
in this situation. On the other hand, the picture of a nearly simultaneous 
break-up in some freeze-out volume seems more justified in this case. 
Indeed, the time scales of less than 100 fm/c are extracted for 
multifragmentation reactions from experimental data 
\cite{beaulieu,karnaukhov}. 
Sophisticated dynamical calculations have also shown that a nearly 
simultaneous break-up into many fragments is the only possible way for 
the evolution of highly-excited systems, e.g. \cite{NMD,XXX}. 
Theoretical arguments in favor of a simultaneous break-up follows also 
from the Hartree-Fock and Thomas-Fermi calculations which predict that 
the compound nucleus will be unstable at high temperatures \cite{HFTF}. 

There exist several analyses of experimental data, which reject the 
binary decay mechanism of fragment production 
via sequential evaporation from a compound nucleus, at high excitation energy. 
For example, this follows from the fact that a popular 
sequential GEMINI code cannot describe the multifragmentation data 
\cite{hubele,Deses,napolit}. We believe that a formal reason of this 
failure is that the evaporation approaches always predict larger 
probabilities for emission of light particles (in particular, neutrons) 
than for intermediate mass fragments (IMFs). 
We mention also attempts to extend the compound nucleus picture by 
including its expansion within the harmonic-interaction Fermi gas (HIFGM) 
model \cite{toke2}, and within the expanding emitting source (EES) model 
\cite{EES}. However, these models have the same theoretical problem with 
short emission times. 
Unfortunately, the EES model has never been compared with 
multifragmentation experiment in 
a comprehensive way since it is limited by considering emission of 
light IMFs with charges $Z\loo 10$ only. 

As was shown already in early statistical model calculations, see e.g. 
\cite{Botvina85}, the entropy of the compound nucleus dominates over 
entropies of multifragmentation channels at low energies, but this trend 
reverses at high excitation energies. This means that the evaporation/fission 
based models can only be 
used at excitation energies below the multifragmentation 
threshold, $E_{th}$= 2--4 MeV/nucleon, but at higher excitations a 
simultaneous emission must be a preferable assumption. 
Close to the onset of multifragmentation 
the most probable decay channels contain one (compound-like), or two 
(fission-like) fragments, 
and a few small fragments. With increasing excitation 
energy the break-up into several IMFs becomes more probable, 
and at very high excitation energies the decay channels with nucleons and 
lightest fragments (vaporization) dominate. Such evolution of nuclear 
decay mechanisms is predicted by all statistical models. 

\vspace{0.5cm}

{\bf 3.2 Statistical models of multifragmentation}\\

Main concepts of the statistical approach to nuclear multifragmentation 
have been formulated in 80-s by Randrup et al. \cite{Randrup}, 
Gross et al. (MMMC) \cite{Gross0}, and Bondorf et al. (SMM) 
\cite{Botvina85,SMM0,Botvina87}. 
This approach is based on the assumption that the relative probabilities of 
different break-up channels are determined by their statistical weights, 
which include contributions of phase space (spatial and momentum) factors 
and level density of 
internal excitations of fragments. Different versions of the model differ 
in details of description of individual fragments, Coulomb interaction and 
choice of statistical ensembles (grand-canonical, canonical, or 
microcanonical). Usually, all these details do not affect significantly 
qualitative features of the statistical break-up. For example, the 
differences in ensembles can hardly be seen in fragment distributions at high 
excitation energies \cite{SMM}, unless the observables are selected in a very 
special way. As was later demonstrated in experiments of many groups: ALADIN 
\cite{ALADIN}, EOS \cite{EOS}, ISIS \cite{ISIS}, Miniball-Multics \cite{MSU}, 
INDRA \cite{INDRA}, FAZA \cite{FAZA}, NIMROD \cite{NIMROD} and others, 
equilibrated sources are indeed formed in nuclear reactions, and statistical 
models are very successful in describing the fragment production from them. 
This proves that the multifragmentation process to a large extend is 
controlled by the available phase space including internal excitations 
of fragments. 
Furthermore, systematic studies of such highly excited systems have brought 
important information about a liquid-gas phase transition in finite nuclear 
systems \cite{Pochodzala,Dagostino2}. 

The success of first statistical models has stimulated appearance of their 
new versions in next decades. The models MMM \cite{MMM} and ISMM 
\cite{ISMM} are based on the same principles and use the same methods 
with small modifications. 
In the SIMON code \cite{durand}, fragments evaporated from the compound 
nucleus are placed in a common volume in order to simulate a simultaneous 
break-up. 
There were also developments of the original models: SMM \cite{Botvina01}, 
and MMMC \cite{lefevre04}, bringing some improvements seen as necessary 
from the analysis of experimental data. 
An interesting mathematical development has been made in 
refs. \cite{DasGupta}, where a canonical version of the SMM 
with simple partition weights was exactly analytically resolved by using 
recursive relations for the partition sum. 
Most models use the Boltzmann statistics, since the number of particular 
fragments in the freeze-out volume is typically of order 1. 
Calculations of ref. \cite{Parvan} have demonstrated that the quantum 
statistical effects do not play a role for all species, but nucleons at 
excitation energies and entropies characterizing multifragmentation. 
The same conclusion has been made in ref. \cite{ZPHYS} by direct 
comparisons of SMM with a quantum statistical model (QSM) \cite{QSM}. 

As a rule, all above mentioned statistical models give very 
similar results concerning description of mean characteristics of 
multifragmentation. For example, description of ALADIN experiments requires 
ensembles of emitting sources which in SMM, MMMC and MMM models differ 
within 10\% of their masses and excitation energies \cite{ALADIN,Xi,MMM00}. 
Such an uncertainty is of the same magnitude as the precision of most 
experimental data. One can see some differences between the models only 
in more sophisticated observables. For example, the isotope properties 
of produced fragments, especially the isoscaling observables, 
may allow for better discrimination between different approaches, 
as well as between parameters within a specific model \cite{tsang,traut}. 

\vspace{0.5cm}

{\bf 3.3 Fragment formation and freeze-out volume}\\

In a simplified consideration, all simultaneously produced fragments are 
placed within a fixed freeze-out volume. It is assumed that nuclear 
interactions between the fragments cease at this point, and at later time 
fragments propagate independently in the mutual Coulomb field. 
In fact, there is a deep physical idea behind this simple picture. 
During the fragment formation the nucleons move in a common mean field, 
and experience stochastic collisions. When collisions practically cease, 
the relatively cold group of nucleons get trapped by the local mean field 
and form fragments \cite{mishust95}. 
It is assumed that there exists a certain point in the space-time evolution,
which is crucial for the final fate of the system. 
This is a so-called 'saddle' point, and the freeze-out volume provides 
a space for the 'saddle' point configurations. 
According to the statistical approach, the probabilities of the fragment 
partitions are determined by their statistical weights at the 'saddle' point. 
Actually, the nuclear interactions between fragments may not cease 
completely after the 'saddle' point, however, they do not change the 
fragment partitions which have been decided at this point. 
Only when the system reaches the 'scission' point the contact between the 
fragments is finally disrupted. 
This picture may be justified by the analogy with nuclear fission, 
where the existence of 'saddle' and 'scission' points is commonly accepted. 

In most statistical models one assumes that 'saddle' and 'scission' points 
coincide and the statistical weight is characterized by a single freeze-out 
volume. On the other hand, one should distinguish the full geometrical volume 
and a so-called 'free' volume, which is available for the fragment 
translational motion in coordinate space. 
Due to the final size of fragments and their mutual interaction this free 
volume is smaller than the physical freeze-out volume, at least, by the 
proper volume of all produced fragments. 
This "excluded volume" can be included in statistical models with different 
prescriptions, which, however, must respect the conservation laws 
\cite{botmis}. 
In the SMM there are two distinct parameters which control the free volume 
and the freeze-out volume. In some respects these two different volumes are 
introduced similar to 'saddle' and 'scission' points discussed above. 
In principal, the different volumes should be extracted from analysis of 
experimental data \cite{EOS,karnaukhov1}. 
Since the entropy associated with the translational motion is typically 
much smaller than the entropy associated with the internal excitation of 
fragments, uncertainties in the determination of the free volume do not 
affect significantly the model predictions, especially in the case of 
break-up into few fragments. 

There are several schematic views of how the fragments are positioned in 
coordinate space. 
The most popular picture assumes expansion of uniform nuclear 
matter to the freeze-out volume, accompanied by its 'cracking' and 
fragment formation. 
However, this picture is more appropriate for the processes with a large 
excitation energy and flow, and corresponds to the transition of the 
nucleon 'gas' to the 
'liquid' drops by cooling during the expansion. This picture can not 
be applied at energies close to the multifragmentation threshold, 
since they are not sufficient for essential uniform expansion 
of the nucleus. 
At $E^*\goo E_{th}$ the picture of a simultaneous 'fission' into several 
fragments seems more appropriate. 
One should bear in mind that for statistical description it is not 
important how the system has evolved toward the 'saddle' point. 
The only assumption in this case is that the phase space and level 
density factors dominate over the transition matrix elements. 
This explains why different models are rather consistent with each 
other irrespective of the way how the fragment positioning is made. 

The average density which corresponds to the freeze-out volume is usually 
taken in the range between 1/3 and 1/10 of the normal nuclear density 
$\rho_0\approx$ 0.15 fm$^{-3}$. In the case of thermal multifragmentation the 
freeze-out density can be reliably estimated from experimental data on 
fragment velocities since they to 80-90\% are determined by the Coulomb 
acceleration after the break-up. 
The experimental analyses of the kinetic energies, angle- 
and velocity-correlations of the fragments indeed point to values of 
(0.1-0.4) $\rho_0$ \cite{aladin1,viola}. 

\vspace{1.0cm}

{\bf 3.4 Fragments in the statistical approach}\\

Another important concept refers to 'primary fragments', i.e., the fragments 
which are produced in the freeze-out volume. Properties of these 
fragments essentially determine statistical weights of partitions. 
The simplest approximation is to use the masses (or binding energies) 
of the nuclei from the nuclear data tables referring to cold isolated 
nuclei, for example, as it is done in MMMC, or in ISMM. In order  
to calculate the contribution of fragment's internal excitations to the 
statistical weight one should introduce additional assumptions concerning 
their level densities. 
For example, the MMMC prescription is 1) to limit the internal excitation of 
fragments by particle stable levels only (this leads to relatively cold 
fragments), and 2) to include in the statistical weight the contribution of 
secondary neutrons, which are assumed to be evaporated 
instantaneously from primary fragments in the freeze-out volume \cite{Gross}. 
Randrup at al. \cite{Randrup} and MMM \cite{MMM} use a Fermi gas type 
approximation with a cut off at high temperatures. 
In the ISMM this is done via level density expressions motivated by empirical 
information for isolated nuclei \cite{ISMM}. 
However, as clear from the previous discussion, the approximations used for 
isolated nuclei may not be true in the freeze-out volume since the fragments 
can still interact and, therefore, have modified properties. 
For example, as was noted long ago, the neutron content of primary fragments 
can be changed due to 
reduced Coulomb interaction in the hot environment of nucleons and other 
fragments \cite{Botvina87,Botvina85,Botvina01}. 

In order to include possible in-medium effects, the SMM has adopted a 
liquid-drop description of individual fragments ($A>$4) extended for 
the case of finite temperatures and densities \cite{SMM}. 
Smaller clusters are considered as elementary particles. 
At low excitation energies this description corresponds to known 
properties of cold nuclei, but it is generalized for the 
consideration of highly excited nuclei in medium. 
The parameters of the liquid drop description change as a result of 
interactions between the fragments leading, in particular, to 
modifications of bulk, surface and Coulomb terms. 
These parameters can be evaluated from the analysis of experimental data. 
As discussed in ref. \cite{traut} possible changes in symmetry energy 
of hot fragments can be extracted from the isoscaling data. The experimental 
evidences have been found that the symmetry energy of hot fragments in the 
freeze-out volume decreases noticeably as compared with cold nuclei 
\cite{LeFevre,Shetty05}. 

We emphasize that in-medium modification of fragment properties  is a natural 
way to include interaction between fragments within the statistical approach. 
Recently, an attempt has been made \cite{Bonasera} to 
consider the evolution of the fragments after freeze-out 
within the framework of a dynamical model with explicit inclusion of nuclear 
interactions. As reported, this interaction results in a fusion 
(recombination) of primary fragments, and thus modifies the fragment 
partitions. However, the 
dynamical fragment formation after the "statistical freeze-out" leads to 
violation of fundamental assumptions of the statistical approach, such as the 
ergodicity and detail balance principles. 
Generally, an application of a time-dependent approach (dynamics) to a 
statistical ensemble would be a controversial operation since, according 
to the ergodicity principle, the time average over microscopic configurations 
must be equivalent to the ensemble average. The dynamical consideration 
may be justified only for the long-range Coulomb forces influencing 
fragments' motion after their formation. 
Therefore, results of ref. \cite{Bonasera} are misleading and 
cannot be considered as an improvement of the statistical approach. 

\vspace{0.5cm}

{\bf 3.5 Influence of flow on fragment formation}\\

As was established experimentally, an 'ideal' picture of 
thermal multifragmentation begins to fail at excitation energies 
of about 5--6 MeV/nucleon \cite{lefort}. At higher 
excitations a part of the energy goes into a collective kinetic 
energy of the produced fragments, without thermalization. This energy is 
defined as the flow energy, and its share depends on the kind of reaction. 
For example, at thermal excitation energy of $E^*\approx$6 MeV/nucleon, the 
additional flow energy is 
around 0.2 MeV/nucleon in hadron-induced reactions, and it is 
around 1.0 MeV/nucleon in central heavy-ion collisions around Fermi energy. 
Since a dynamical flow itself can break matter into pieces, it is necessary 
to understand limits of the statistical description in the case of a strong 
flow. 

This problem was addressed in the number of works within dynamical and 
lattice-gas models \cite{NMD,DasGupta2,Chomaz1,mishust}. Their conclusion 
is that a flow does not change statistical model predictions, if its energy 
is essentially smaller than the thermal energy. 
This justifies a receipt often used in statistical models, when 
the flow energy is included by increasing the velocities of fragments in 
the freeze-out volume according to the flow velocity profile \cite{SMM}. 
This is in agreement with many experimental analyses. However, statistical 
models work surprisingly well even when the flow energy is comparable with 
the thermal energy, or even higher \cite{lefevre04,neubert}. 
This observation requires additional study.

\vspace{0.5cm}

{\bf 3.6  Nuclear liquid-gas phase transition within statistical models}\\ 

Many statistical models have demonstrated that multifragmentation is a kind 
of a phase transition in highly excited nuclear systems. 
In the SMM a link to the liquid-gas phase transition is especially strong. 
In particular, the surface energy of hot primary fragments is 
parametrized in such a way that it vanishes at a certain critical 
temperature. The SMM has predicted distinctive features of this phase 
transition in finite nuclei, such as the plateau-like anomaly in the 
caloric curve \cite{SMM0,SMM}, 
which have been later observed in experiments \cite{Pochodzala,Natowitz}. 
Many other manifestations of the phase transition, such as large 
fluctuations and bimodality \cite{ALADIN,Dagostino2,nihal}, critical behavior 
and even values of critical exponents \cite{Dagostino2,Srivastava}, have been 
investigated within this model. The experimental data are usually in agreement 
with the predictions. 

Nevertheless, the properties of this phase transition are not yet fully 
understood. 
The critical behavior observed in experimental data can also be explained 
within a percolation model \cite{percolation}, or a Fisher's droplet model 
\cite{moretto2}, which correspond to a second order phase transition in 
the vicinity of the critical point. We must note, however, that the 
finiteness of the systems under investigation plays a crucial role. 
To connect this anomalous behavior with a real phase transition one should 
study it in a thermodynamical limit. Within the SMM this was done in 
ref. \cite{bugaev}, where multifragmentation of an equilibrated system 
was identified as a first order phase transition. 
The mixed phase in this case consists of an infinite liquid condensate and 
gas of nuclear fragments of all masses. In a finite system this mixed phase 
corresponds to U-shaped fragment distributions with a heaviest 
fragment representing the liquid phase. 
Thus one can connect multifragmentation of finite nuclei 
with the fragmentation of a very big system. 
This is important for the application of statistical models in astrophysical 
environments (neutron stars, supernova explosions), where nuclear 
statistical equilibrium can also be expected \cite{Botvina04}. 

\vspace{0.5cm}

{\bf 3.7 Relation between statistical and dynamical descriptions}\\

One of the problems, which is highly debated now, is if dynamical models 
alone can describe (at least qualitatively) the same evolutionary scenario 
leading to equilibration and multifragmentation as assumed by 
statistical models. In other words, is it 
possible to use only a "universal" dynamical description, 
instead of subdividing the process into dynamical and statistical stages? 
Some dynamical approaches try to reach this goal starting from 'first 
principles' like Fermionic Molecular Dynamics (FMD) \cite{Feldmeier}, 
or Antisymmetrized Molecular Dynamics (AMD) \cite{Ono}. Other approaches, 
like QMD \cite{dynmodels,larionov}, NMD \cite{NMD}, or BNV \cite{ditoro} 
use classical equations including two-body collisions and some elements 
of stochasticity. 
In all cases dynamical simulations are more complicated and time-consuming 
as compared with statistical models. This is why full calculations, 
e.g. with FMD and AMD models, can only be done for relatively light systems. 
By using simplified receipts, like a coalescence for final fragment 
definition in AMD, one may reduce the computing time, but it still 
remains rather long. 
This prevents from including these codes into practical transport 
calculations in extended complex medium. 

One should bear in mind that the statistical and dynamical approaches are 
derived from different physical principles. The time-dependent dynamical 
approaches are based on Hamiltonian dynamics (the principle of minimal 
action), whereas the statistical models employ the principle of uniform 
population of the phase space. Actually, these two principles are not 
easily reducible to each other, and they represent complementary methods 
for describing the physical reality. There are numerous studies of the 
phase space population with dynamical models (see e.g. \cite{NMD1}), 
which, however, have never shown the uniform population in the limit of 
long times. Therefore, a decision of using statistical or dynamical 
approaches for description of nuclear multifragmentation should be made 
after careful examination of the degree of equilibration expected in 
particular cases, and it can be only justified by comparison with experiment. 

There is still a large difference in details between 'statistical' and 
'dynamical' description of individual fragments as finite quantum systems. 
Usually, the realistic description of clustering is difficult to achieve in 
dynamical models dealing with 
individual nucleons, but it is easily done in statistical models, 
considering nuclear fragments as independent degrees of freedom. 
In case of equilibrated sources predictions of statistical models are 
usually in better agreement with experimental data. 
A most striking example concerns isospin characteristics. 
Dynamical models predict decreasing neutron-richness 
of intermediate mass fragments in collisions of neutron-rich 
nuclei with increasing centrality \cite{ditoro}, 
i.e., with increasing excitation energy. However, experimental data 
demonstrate an opposite trend both in the 'neck region' \cite{Xu} and 
in the equilibrated sources \cite{Milazzo}. 
On the other hand, these trends can easily be explained in 
the framework of the statistical model \cite{Botvina01}. Dynamical models 
are not very successful in describing isoscaling observables 
(e.g., the slope coefficients) \cite{ditoro2}, 
while they are naturally explained within statistical approaches 
\cite{tsang,traut}.

\vspace{0.5cm}

{\bf 4. Deexcitation and propagation of hot primary fragments}\\

After production in the freeze-out volume primary fragments will 
propagate in mutual Coulomb field and undergo deexcitation. 
It is usually assumed that the long-range Coulomb force, 
which has participated only partly in the fragment formation, is fully 
responsible for the post--freeze-out acceleration of the fragments. 
All statistical models 
solve classical Newton equations, taking into account the initial 
positions of fragments inside the freeze-out volume and their thermal 
velocities. At this stage the collective flow of fragments 
can also be taken into consideration. 

The hot fragments will lose excitation in the course of their 
propagation to detectors. There are different secondary deexcitation codes 
used in multifragmentation studies. The standard fission-evaporation 
and Fermi-break-up codes described in \cite{Botvina87} were used in SMM 
\cite{SMM} and MMM \cite{MMM00}. 
Another procedure, which includes GEMINI for deexcitation of big fragments, 
was adopted in the ISMM \cite{ISMM}. In the MMMC \cite{Gross} a schematic 
model was used which takes into account only early emission of secondary 
neutrons. 
Apparently, this oversimplification is responsible for deviations of the 
MMMC predictions from other models in description of correlations between 
neutrons and charged particles \cite{Toke}. 
It should be emphasized that most deexcitation models are based on 
properties of cold isolated nuclei, known from experiments at low energy. 
At present there is a need in more advanced models, which take into 
account possible in-medium modifications of primary fragments in the 
freeze-out volume, e.g., changing their symmetry and surface energy. 
An example of such a model is presented in ref. \cite{nihal}. 

The deexcitation process depends strongly on nuclear content of the 
primary fragments. For example, in the SMM at $E^*$ slightly above 
$E_{th}$ almost all nucleons are contained in fragments, 
the fraction of free nucleons is negligible. This shows 
an analogy with the fission process. As a result the neutron 
content of primary fragments is nearly the same as in the initial source. 
The outcome of deexcitation depends on the actual code used. 
Generally, in realistic statistical models most neutrons come from the 
secondary deexcitation stage, for example, more than 90\% in the SMM. 
If one takes into account a reduction of the symmetry energy 
of primary fragments, and includes its restoration in the course of 
deexcitation, the neutron richness of cold final fragments will be larger 
than predicted by standard codes \cite{nihal}.

\vspace{0.5cm}

{\bf  5. Conclusions}\\

We believe that statistical models suit very well for description 
of such a complicated many-body process as nuclear multifragmentation. 
If a thermalized source can be recognized in a nuclear 
reaction, the main features of multiple fragment production can 
be well described within the statistical approach. 
The success of statistical models in describing a broad range of 
experimental data gives us confidence that this approach will be used and 
further developed in the future. 
We especially stress two main achievements of statistical models 
in theory of nuclear reactions: first, a clear understanding has been 
reached that sequential decay via compound nucleus must give a way to 
nearly simultaneous break-up of nuclei at high excitation energies; 
and, second, the character of this change can be interpreted 
as a liquid-gas type phase transition in finite nuclear systems.

The results obtained in the nuclear multifragmentation studies 
can be applied in several other fields. 
First, the mathematical methods of the statistical multifragmentation can be 
used for developing thermodynamics of finite systems \cite{Gross1,Chomaz}. 
These studies were stimulated by recent observation of extremely large 
fluctuations of energy of produced fragments, which can be interpreted 
as the negative heat capacity \cite{dagostino00}. 
At this point one can see links with 
cluster physics and condensed matter physics \cite{Gross1}. 
These methods might also be useful for investigating possible phase 
transitions from hadronic matter to quark-gluon plasma in relativistic 
heavy-ion collisions. 

Another conclusion is related to the fact that the multifragmentation 
channels take as much as 10-15\% of the total cross section in high-energy 
hadron-nucleus reactions, and about twice more in high-energy nucleus-nucleus 
collisions. Moreover, multifragmentation reactions are responsible for 
production of some specific isotopes. 
Importance of multifragmentation reactions is now widely recognized, and in 
recent years the interest to them has risen in several domains of research.
Indeed, practical calculations of fragment production and transport in 
complex medium are needed for: nuclear waste transmutation (environment 
protection), electro-nuclear breeding 
(new methods of energy production), proton and ion therapy (medical 
applications), radiation protection of space missions (space research). 
Until recently, only evaporation and fission codes have been used for 
describing the nuclear deexcitation. 
We believe that the state-of-the-art today requires inclusion of 
multifragmentation reactions in these calculations. 
The SMM is especially suitable for this purpose because of its 
multifunctional code structure: Besides the multifragmentation channels 
it includes also compound nucleus decays via evaporation and 
fission, and takes into account competition between all channels. 
Encouraging attempts to construct hybrid models, 
combining dynamical and statistical approaches, were undertaken in 
refs. \cite{Botvina90,buusmm,qmdsmm}. The hybrid models are quite successful 
in describing data, including correlation observables 
between dynamical and statistical stages \cite{SMM,volant,turzo}. 
Several multi-purpose codes, like GEANT4 \cite{GEANT4}, have been developed 
to describe transport of hadrons and ions in extended medium. 
The SMM was included in this code as important part responsible for 
fragment production. 

It is important that nuclear multifragmentation reactions 
allow for experimental determination of in-medium modifications of 
hot nuclei/fragments in hot and dense environment. 
This opens the unique possibility for investigating 
the phase diagram of nuclear matter at temperatures $T \approx 3-8$ MeV 
and densities around $\rho \approx 0.1-0.3 \rho_0$, which are expected in the 
freeze-out volume. These studies are complementary to the
previous studies of isolated nuclei existing in the matter with terrestrial
densities, and at low temperatures, $T < 1-2$ MeV. 
The experimental information on properties of hot nuclei in dense 
surrounding is crucial for construction of a reliable equation of state of 
stellar matter and modeling nuclear composition in supernovae 
\cite{Botvina04}. 
This shows that studying the multifragmentation reactions in the laboratory 
is important for understanding how heavy elements were synthesized in 
the Universe. 

We thank S. Das Gupta for careful reading the manuscript and positive 
comments. Also, we thank participants of the WCI workshops for stimulating 
discussions, and organizers, for hospitality and support. 
This work was supported in part by the grant RFFR 05-02-04013 (Russia).


\begin{thebibliography}{99}

\bibitem{Bohr} N. Bohr,  Nature, {\bf 137}, 344 (1936).

\bibitem{Gross} D.H.E. Gross, Rep. Progr. Phys. {\bf 53}, 605 (1990).

\bibitem{SMM} J.P. Bondorf, A.S. Botvina, A.S. Iljinov, I.N. Mishustin and
K. Sneppen,  Phys. Rep. {\bf 257}, 133 (1995).

\bibitem{DasGupta1} S.Das Gupta, A.Z. Mekjian, and B. Tsang, 
Adv. Nucl. Phys. {\bf 26}, 89 (2001). 

\bibitem{Botvina90} A.S. Botvina, A.S. Iljinov, and I.N. Mishustin, 
 Nucl. Phys. {\bf A507}, 649 (1990).

\bibitem{dynmodels} H. Stoecker and W. Greiner, Phys. Rep. {\bf 137}, 277 
(1986). J. Aichelin et al., Phys. Rev. {\bf C37}, 2451 (1988). 
W. Bauer et al., Ann. Rev. Nucl. Part. Sci. {\bf 42}, 77 (1992). 
B.A. Li, Phys. Rev. {\bf C47}, 693 (1993).
J.Konopka et al., Prog. Part. Nucl. Phys., {\bf 30}, 301 (1993).
M. Colonna et al., Nucl. Phys. {\bf A589}, 160 (1995). 
C. Fuchs and H.H. Wolter, Nucl. Phys. {\bf A589}, 732 (1995). 

\bibitem{buusmm} H.W. Barz et al., Nucl. Phys. {\bf A561}, 466 (1993).

\bibitem{Bondorf94} J.P. Bondorf, A.S. Botvina, I.N. Mishustin, 
and S.R. Souza, Phys. Rev. Lett. {\bf 73}, 628 (1994).

\bibitem{larionov} A.S. Botvina, A.B. Larionov, and I.N. Mishustin, 
 Phys. Atom. Nucl. {\bf 58}, 1703 (1995). 

\bibitem{NMD} J.P. Bondorf, O. Friedrichsen, D. Idier, and I.N. Mishustin, 
Nucl. Phys. {\bf A624}, 706 (1997). 

\bibitem{Moretto} L.G. Moretto, Nucl. Phys. {\bf A247}, 211 (1975).

\bibitem{Botvina87} A.S. Botvina et al., Nucl. Phys. {\bf A475}, 663 (1987).

\bibitem{charity0} R.J. Charity et al., Nucl. Phys. {\bf A483}, 371 (1988).

\bibitem{charity} R.J. Charity, Phys. Rev. {\bf C61}, 054614 (2000).

\bibitem{jandel} M. Jandel et al., J. Phys. {\bf G: 31}, 29 (2005).

\bibitem{beaulieu} L. Beaulieu et al., Phys. Rev. Lett. {\bf 84}, 5971 (2000).

\bibitem{karnaukhov} V.A. Karnaukhov et al., Phys. Atom. Nucl. {\bf 66}, 
1242 (2003).

\bibitem{XXX} S.Das Gupta et al., Phys. Rev. {\bf C35}, 556 (1987). 
B. Strack, Phys. Rev. {\bf C35}, 691 (1987). 
D.H. Boal and J.N. Gloshi, Phys. Rev. {\bf C37}, 91 (1988).

\bibitem{HFTF} P. Bonche, S. Levit, and D. Vautherin, 
Nucl. Phys. {\bf A436}, 265 (1985). E. Suraud, Nucl. Phys. {\bf A462}, 
109 (1987). 

\bibitem{hubele} J. Hubele et al., Phys. Rev. {\bf C46}, R1577 (1992). 

\bibitem{Deses} P. Desesquelles et al., Nucl. Phys. {\bf A604}, 183 (1996). 

\bibitem{napolit} P. Napolitani et al., Phys. Rev. {\bf C70}, 054607 (2004). 

\bibitem{toke2} J. Toke et al., Phys. Rev. {\bf C72}, R031601 (2005). 

\bibitem{EES} W.A. Friedman, Phys. Rev. {\bf C42}, 667 (1990). 

\bibitem{Botvina85} A.S. Botvina, A.S. Iljinov, and I.N. Mishustin,  
Sov. J. Nucl. Phys. {\bf 42}, 712 (1985).

\bibitem{Randrup} J. Randrup and S.E. Koonin, Nucl. Phys. {\bf A356}, 
223 (1981). 
S.E. Koonin and J. Randrup, Nucl. Phys. {\bf A471},
355c (1987). J. Randrup, Comp. Phys. Comm. {\bf 77}, 153 (1993). 

\bibitem{Gross0} D.H.E. Gross et al., Z. Phys. {\bf A309}, 41 (1982). 
X.Z. Zhang, D.H.E. Gross, S. Xu, and Y.M. Zheng, Nucl. 
Phys. {\bf A461}, 641 (1987).

\bibitem{SMM0} J.P. Bondorf, R. Donangelo, I.N. Mishustin, C.J. Pethick, 
H.Schulz, and K. Sneppen, Nucl. Phys. {\bf A443}, 321 (1985). 
J.P. Bondorf, R. Donangelo, I.N. Mishustin, and H. Schulz, Nucl. Phys. 
{\bf A444}, 460 (1985). 

\bibitem{ALADIN} A.S. Botvina et al.,  Nucl. Phys. {\bf A584},
737 (1995).

\bibitem{EOS} R.P. Scharenberg et al., Phys. Rev. {\bf C64},
054602 (2001).

\bibitem{ISIS} L. Pienkowski et al., Phys. Rev. {\bf C65},
064606 (2002).

\bibitem{MSU} M. D'Agostino, et al., Phys. Lett. {\bf B371}, 175 (1996).

\bibitem{INDRA} N. Bellaize, et al., Nucl. Phys. {\bf A709}, 367 (2002).

\bibitem{FAZA} S.P. Avdeyev et al., Nucl. Phys. {\bf A709}, 392 (2002).

\bibitem{NIMROD} J. Wang et al., Phys. Rev.{\bf C71}, 054608 (2005).

\bibitem{Pochodzala} J. Pochodzalla et al.,  Phys. Rev.
Lett. {\bf 75}, 1040 (1995).

\bibitem{Dagostino2} M. D'Agostino, A.S. Botvina, M. Bruno, A. Bonasera,
J.P. Bondorf, I.N. Mishustin et al., Nucl. Phys. {\bf A650}, 329 (1999).

\bibitem{MMM} Al.H. Raduta, and Ad.R. Raduta, Phys. Rev. {\bf C55}, 1344 
(1997). 

\bibitem{ISMM} M.P. Tan et al., Phys. Rev. {\bf C68}, 034609 (2003). 

\bibitem{durand} D. Durand, Nucl. Phys.  {\bf A541}, 266 (1992).

\bibitem{Botvina01} A.S. Botvina, I.N. Mishustin, Phys. Rev. {\bf C63},
061601(R) (2001).

\bibitem{lefevre04} A.Le Fevre et al., 
Nucl. Phys. {\bf A735}, 219 (2004). 

\bibitem{DasGupta} S.Das Gupta, and A.Z. Mekjian, Phys. Rev. {\bf C57}, 1361 
(1998). 

\bibitem{Parvan} A.S. Parvan et al., Nucl. Phys. {\bf A676}, 409 (2000). 

\bibitem{ZPHYS} A.S. Botvina et al., Z. Phys. {\bf A345}, 297 (1993). 

\bibitem{QSM} D. Hahn, and H. Stoeker, Nucl. Phys. {\bf A476}, 718 (1988). 

\bibitem{Xi} H. Xi et al., Z. Phys. {\bf A359}, 397 (1997). 

\bibitem{MMM00} Al.H. Raduta, and Ad.R. Raduta, Phys. Rev. {\bf C61}, 034611 
(2000). 

\bibitem{tsang} M.B. Tsang et al., Phys. Rev. {\bf C64}, 054615 (2001).

\bibitem{traut} A.S. Botvina, O.V. Lozhkin and W. Trautmann,
             Phys. Rev. {\bf C65}, 044610 (2002).

\bibitem{mishust95} J.P. Bondorf, D. Idier, and I.N. Mishustin, Phys. Lett. 
{\bf B359}, 261 (1995). 

\bibitem{botmis} A.S. Botvina and I.N. Mishustin, Phys. Rev. Lett. 
{\bf 90}, 179201 (2003). 

\bibitem{karnaukhov1} V.A. Karnaukhov et al., Phys. Rev. {\bf C70}, 
041601 (2004). 

\bibitem{aladin1} S. Fritz et al., Phys. Lett. {\bf B461}, 315 (1999). 

\bibitem{viola} V.E. Viola et al., Phys. Rev. Lett. {\bf 93}, 
132701 (2004). 

\bibitem{LeFevre} A. Le Fevre et al., Phys. Rev. Lett. {\bf 94}
162701 (2005).

\bibitem{Shetty05} D.V. Shetty et al., Phys. Rev. {\bf C71},
024602 (2005).

\bibitem{Bonasera} S.K. Samaddar, J.N. De, and A. Bonasera, 
Phys. Rev. {\bf C71}, 011601, 2005. 

\bibitem{lefort} T. Lefort et al., Phys. Rev. {\bf C62}, 031604 (2000). 

\bibitem{DasGupta2} C.B. Das, L. Shi, and S.Das Gupta, Phys. Rev. 
{\bf C70}, 064610 (2004).

\bibitem{Chomaz1} F. Gulminelli adn Ph. Chomaz, Nucl. Phys. {\bf A734}, 
581 (2004).

\bibitem{mishust} I.N. Mishustin, AIP Conf. Proc. {\bf 512}, 308 (2000).

\bibitem{neubert} W. Neubert, and A.S. Botvina, Eur. Phys. J. 
{\bf A17}, 559 (2003). 

\bibitem{Natowitz} J.B. Natowitz et al., 
Phys. Rev. {\bf C65}, 034618 (2002). 

\bibitem{nihal} N. Buyukcizmeci et al., Eur. Phys. J. {\bf A25}, 57 (2005).

\bibitem{Srivastava} B.K. Srivastava et al., 
Phys. Rev. {\bf C65}, 054617 (2002). 

\bibitem{moretto2} J.B. Elliott et al.,  Phys. Rev. Lett. {\bf 88},
042701 (2002).

\bibitem{percolation} W. Bauer and A.S. Botvina,  Phys. Rev. {\bf C52}, 
R1760 (1995). M. Kleine Berkenbusch et al., Phys. Rev. Lett. {\bf 88}, 
022701 (2002). 

\bibitem{bugaev} K.A. Bugaev et al.,  Phys. Rev. {\bf C62},
044320 (2000).

\bibitem{Botvina04} A.S. Botvina and I.N. Mishustin, Phys. Lett.
{\bf B584}, 233 (2004).

\bibitem{Feldmeier} H. Feldmeier, Nucl. Phys. {\bf A681}, 398c (2001). 

\bibitem{Ono} A. Ono and H. Horiuchi, Phys. Rev. {\bf C53}, 2958 (1996).

\bibitem{NMD1} J.P. Bondorf, H. Feldmeier, I.N. Mishustin, and G. Neergaard, 
Phys. Rev. {\bf C65}, 017601 (2002). 

\bibitem{ditoro} V. Baran et al., Nucl. Phys. {\bf A703}, 603 (2002).

\bibitem{Xu} H. Xu et al., Phys. Rev. {\bf C65}, 061602 (2002).

\bibitem{Milazzo} P.M. Milazzo et al., Phys. Rev. {\bf C62}, 041602 (2000).

\bibitem{ditoro2} T.X. Liu et al., Phys. Rev. {\bf C69}, 014603 (2004).

\bibitem{Toke} J. Toke et al., Phys. Rev. {\bf C63}, 024604 (2001).

\bibitem{Gross1} D.H.E. Gross, Phys. Rep. {\bf 279}, 119 (1997).

\bibitem{Chomaz} Ph. Chomaz, F. Gulminelli, and V. Duflot, 
Phys. Rev. {\bf E64}, 046114 (2001). 

\bibitem{dagostino00} M. D'Agostino et al., Phys. Lett. {\bf B473}, 219 
(2000). 

\bibitem{qmdsmm} T.C. Sangster et al., Phys. Rev. {\bf C46}, 1404 (1992).

\bibitem{volant} C. Volant et al., Nucl. Phys. {\bf A734}, 545 (2004). 

\bibitem{turzo} K. Turzo et al., Eur. Phys. J. 
{\bf A21}, 293 (2004). 

\bibitem{GEANT4} S. Agostinelli et al.,  NIM {\bf A506}, 250 (2003).

%
%

\end{thebibliography}
\end{document}